\documentclass[12pt]{iopart}
\usepackage{iopams}
\expandafter\let\csname equation*\endcsname\relax
\expandafter\let\csname endequation*\endcsname\relax
\usepackage{amsfonts}
\usepackage{amsmath}
\usepackage{amssymb}
\usepackage{graphicx} 
\usepackage{braket}
\usepackage{enumitem}
\usepackage{xcolor}
\newcommand{\beq}{\begin{equation}}
\newcommand{\eeq}{\end{equation}}

\newcommand{\beqa}{\begin{eqnarray}}
\newcommand{\eeqa}{\end{eqnarray}}
\newcommand{\beqal}{\begin{equation}\begin{aligned}}
\newcommand{\eeqal}{\end{aligned}\end{equation}}

\begin{document}

\title{Ultra-fast two-qubit ion gate using sequences of resonant pulses}

\author{E. Torrontegui$^1$, D. Heinrich$^{2,3}$, M. I. Hussain$^{2,3}$, R. Blatt$^{2,3}$, and J. J. Garc{\'\i}a-Ripoll$^1$}
\address{$^1$ Instituto de F\'{\i}sica Fundamental IFF-CSIC, Calle Serrano 113b, 28006 Madrid, Spain}
\address{$^2$ Institut f\"ur Quantenoptik und Quanteninformation, \"Osterreichische Akademie der Wissenschaften, Technikerstr. 21a, 6020 Innsbruck, Austria}
\address{$^3$ Institut f\"ur Experimentalphysik, Universit\"at Innsbruck, Technikerstr. 25, 6020 Innsbruck, Austria}
\ead{eriktorrontegui@gmail.com}

\begin{abstract}
We propose a new protocol to implement ultra-fast two-qubit phase gates with trapped ions using spin-dependent kicks induced by resonant transitions.
By only optimizing the allocation of the arrival times in a pulse train sequence  the gate is implemented in times faster than the trapping oscillation period $T<2\pi/\omega$.
Such gates allow us to increase the number of gate operations that can be completed within the coherence time of the ion-qubits favoring the development of
scalable quantum computers.

\end{abstract}  	
\maketitle

\section{Introduction}

Trapped ions are one of the most accurate platforms for scalable quantum computation. Many ions can be loaded in Paul traps\ \cite{Porras2004, Zhang2017}, Penning traps\ \cite{Jordan2019} or possibly in other scalable architectures\ \cite{Lekitsch2017,Jain2018}. Within these traps, qubits can be stored in long-lived atomic states, which are individually manipulated using lasers or microwaves to implement high-fidelity single-qubit operations and measurements. Finally, using the vibrational states of the ion crystal mediators, it is possible to implement univeral multiqubit operations, such as the CNOT gate\ \cite{Cirac1995, Schmidt-Kaler2003}, the M{\o}lmer-S{\o}rensen gate \cite{Sorensen1999, Sackett2000}, geometric phase gates\ \cite{Milburn2000, Leibfried2003} or Toffoli operations\ \cite{Toffoli1980, Monz2009}. The actual realization of many of these gates depends on Raman transitions\ \cite{Wolf2016, Kaufmann2017, Ballance2016, Gaebler2016}, with high-fidelity \cite{Gaebler2016, Ballance2016} and excellent coherence properties \cite{Wang2017}. In practice, fidelity and speed of two-qubit gates are still limiting the depth of actual computations, and prevent the development of scalable fault tolerant computation\ \cite{Bermudez2017}.

Those limitations in fidelity and speed are due to the use of highly detuned lasers, with lengthy control procedures and slow dynamics of the vibrational states. There exist faster gates based on faster and stronger acceleration of the ions\ \cite{Garcia-Ripoll2003,Garcia-Ripoll2005,Duan2004,Steane2014}. Already a strong time-dependent optical lattice may result in high-fidelity gates that are shorter than a trap period\ \cite{Schafer2018}, but are still constrained by available detuning, power and the Lamb-Dicke limit\ \cite{Wineland1998}. Another method is to excite an optical transition using picosecond laser pulses. A properly designed pulse train can create an arbitrarily fast two- or multi-qubit gate\ \cite{Garcia-Ripoll2003,Garcia-Ripoll2005}. However, as demonstrated in Ref.\ \cite{Mizrahi2013}, it remains a technical challenge to have a strong momentum kick per pulse---a Raman transition might not provide enough momentum---and to switch directions in the pulse laser---which may induce additional sources of error and decoherence.

In this work we study the realization of fast high-fidelity quantum gates using a train of laser pulses that excite a resonant transition\ \cite{Heinrich2019}. We focus on a simple scenario that only requires pulse-picking from a train of laser pulses with fixed strength and repetition rate. As example, we study a realistic pulsed scheme driving the $4\mathrm{S}_{1/2}\to 4\mathrm{P}_{3/2}$ transition in ${}^{40}\mathrm{Ca}^+$\ \cite{Heinrich2019}. We design the gate protocols with a two-stage global optimization that combines a continuous approximation with a discrete genetic algorithm for fine-tuning the pulse picking. We find many choices of pulses that implement highly entangling gates in a time comparable to the trap frequency, with very weak sensitivity to the pulse arrival time or the temperature of the motional states.

The manuscript is structured as follows: In Sec.\ \ref{gate-theory}, we revisit the theory for implementing phase gates using spin-dependent kicks\ \cite{Cirac1995, Garcia-Ripoll2003, Garcia-Ripoll2005}. Section\ \ref{setup} presents a possible experimental setup and an optimized control protocol based on state-of-the-art kicking and control of trapped ions\ \cite{Heinrich2019}. The results leading to the implementation of ultra-fast two-qubit gates are discussed in Sec.\ \ref{results}. In Sect.\ \ref{errors} we analyze and quantify the main source of errors in the design of such gates. Finally, we present prospective research lines related to this work in Sec.\ \ref{outlook}.
\section{Methods}
\subsection{Geometric phase by state-dependent kicks}
\label{gate-theory}

Consider two ions in a 1D-harmonic potential of frequency $\omega,$ at positions $x_1$ and $x_2$. Using the center-of-mass (c) and stretch-mode (s) coordinates, $x_c=(x_1+x_2)/2$ and $x_s=x_2-x_1$ the free Hamiltonian for this system reads $H_0=\hbar\omega_ca_c^{\dag}a_c+\hbar\omega_sa_s^{\dag}a_s.$ 
Here $\omega_c=\omega$ and $\omega_s=\omega\sqrt{3}$ and $a_{c,s}^{\dag}$ $(a_{c,s})$ are the creation (annihilation) phonon operators for each mode. The ions interact with a laser beam that is resonant with an atomic transition. This interaction is modeled by the effective Hamiltonian
\begin{equation}
\label{Hi}
H_1=\frac{\Omega(t)}{2}[\sigma_1^{\dag}e^{i\hbar k x_1}+\sigma_2^{\dag}e^{i\hbar k x_2}+\mbox{H.c.}].
\end{equation}
The pseudospin ladder operator $\sigma_i^{\dag}$ connects the ground and excited states of the $i$-th ion---in this setup, the $4\mathrm{S}_{1/2}$ and $4\mathrm{P}_{3/2}$ states of ${}^{40}\mathrm{Ca}^+.$ The interaction accounts for processes where the ion absorbs or emits a photon, changing its internal state and also modifying its momentum by $\pm\hbar k.$ The sign of $k$ depends on the direction of the laser and whether the photon is emitted or absorbed. Without loss of generality, we will forego individual addressing and assume that the Rabi frequency $\Omega(t)$ is the same for both ions.

%
\begin{figure}[t]
\begin{center}
\includegraphics[width=0.5\linewidth]{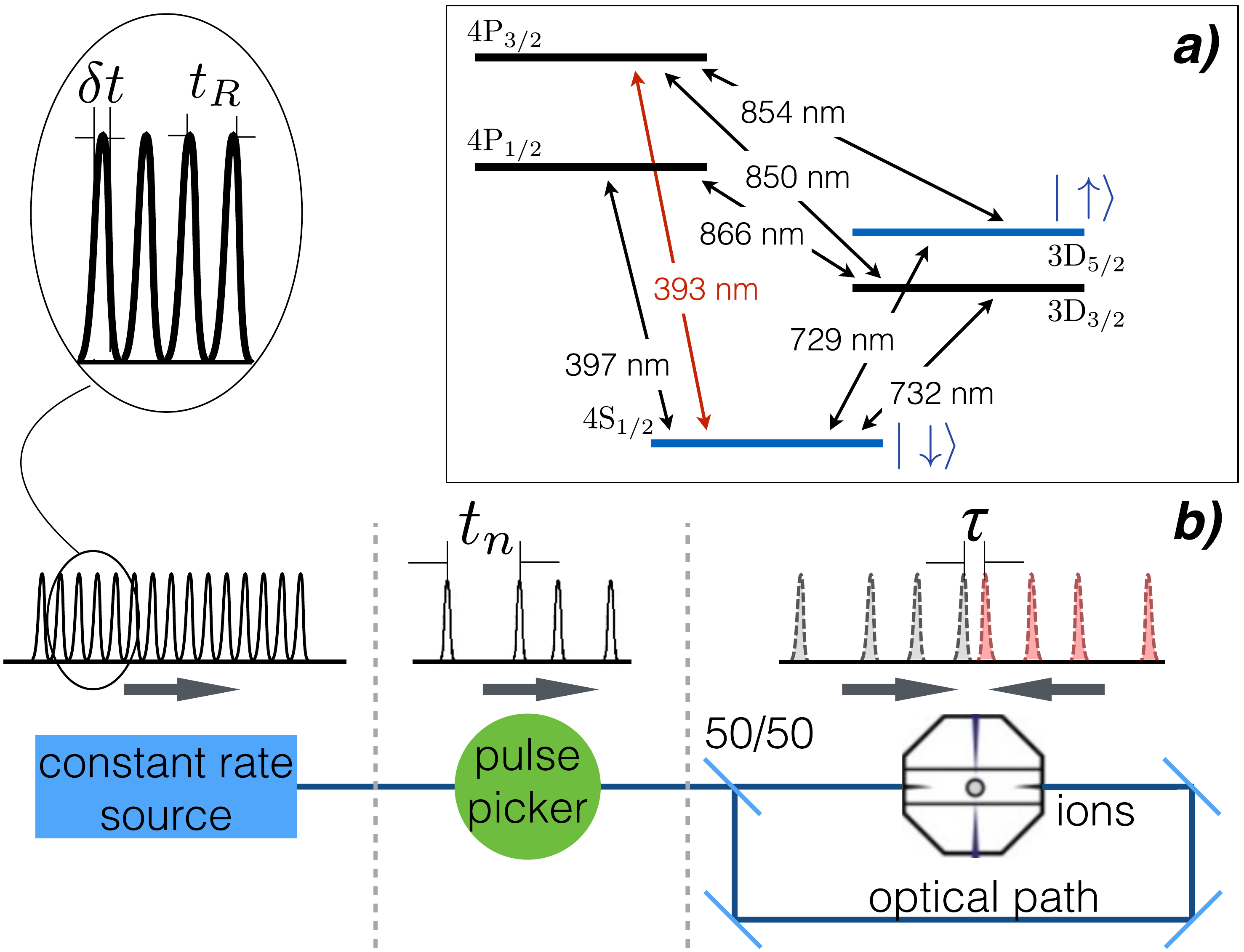}

\caption{{\itshape a)} Relevant levels of a $^{40}\mbox{Ca}^+$ ion. The qubit states $|\downarrow\rangle$ and $|\uparrow\rangle$ correspond
to the $4\mbox{S}_{1/2}$ and $3\mbox{D}_{5/2}$ levels, respectively. A picosecond laser beam of $393.4$ nm, resonant with the 
$4\mbox{S}_{1/2}\leftrightarrow 4\mbox{P}_{3/2}$ transition, imparts
spin-dependent kicks to the ion.
{\itshape b)} (Bottom) Experimental setup. A constant rate source generator creates pulses with an interval $t_R$.
A pulse picker selects the $t_n$ optimal positions and a 50/50 beamsplitter divides each pulse. Both sequences follow a different optical path such that when they arrive at the ion in counter-propagating directions each left-right pulse pair has a relative delay $\tau$ much shorter than the characteristic decay time of the selected transition $t_\gamma.$ (Top) Pulse train at each stage of the experiment, the black arrows represent the motional sense of the pulses. 
 }
\label{fig:levels}
\end{center}
\end{figure}
%
%
Our gate protocols\ \cite{Garcia-Ripoll2003} alternate free evolution $H=H_0$, where the laser is switched off $(\Omega=0),$ with a very fast, pulsed interaction kicking the ion. As shown in Fig.\ \ref{fig:levels}b, we assume pairs of pulses coming from counter-propagating directions. The Rabi frequency $\Omega(t)$ and the duration of each pulse $\delta t$ satisfy $\int_0^{\delta t}\Omega(\tau)d\tau=\pi$ and $\delta t\ll 2\pi/\omega$. The pulses kick the ions, accelerating them along the same direction. In between each pair of kicks, the ions oscillate freely in the trap. The combination of both effects can be modeled analytically. The evolution operator for $N$ pulses is $\mathcal{U}=\mathcal{U}_c\mathcal{U}_s$ with $\mathcal{U}_{c,s}=\prod_{n=1}^NU_{c,s}(t_n,z_n)$ and
\beqal
U_c(t_n,z_n)&=&e^{i\alpha_c^{(n)}(\sigma_1^z+\sigma_2^z)(a_c+a_c^{\dag})}e^{i\omega_ct_na_c^{\dag}a_c} \\
U_s(t_n,z_n)&=&e^{i\alpha_s^{(n)}(\sigma_1^z-\sigma_2^z)(a_s+a_s^{\dag})}e^{i\omega_st_na_s^{\dag}a_s}.
\eeqal
The amplitudes $\alpha_c=\eta z/2^{3/2}$ and $\alpha_s=\alpha_c/3^{1/4}$ depend on the Lamb-Dicke parameter $\eta=\sqrt{\frac{\hbar}{2m\omega}}k.$ The sign $z=\pm 1$ indicates the net orientation of the combined kick. It depends on the relative order of pulses within each pair: $z=+1$ if the first pulse comes from the left and the second from the right, $z=-1$ in the opposite case. In the setup from Fig.\ \ref{fig:levels}b, the sign $z$ is fixed throughout the experiment.
%
\begin{figure}[t]
\begin{center}
\includegraphics[width=0.5\linewidth]{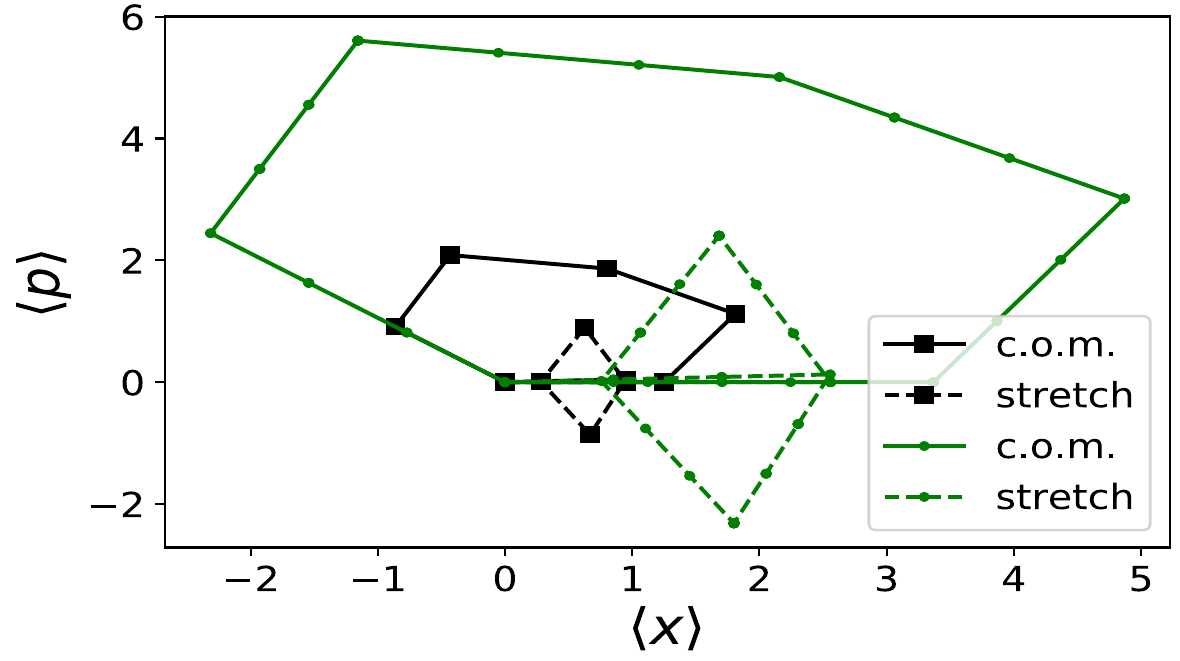}
\caption{Phase space trajectories for the center-of-mass (solid) and stretch mode (dashed) for a pulse sequence with $N=6$ sets of pulses, with $M=1$ pulses (black) and $M=3$ pulses (green) per set, respectively. The trajectories are drawn in the frame of reference that rotates with the frequency of the mode, that is $\braket{a e^{i\omega_{c,s}t}}=\frac{1}{\sqrt{2}}\braket{x_{c,s}}+\frac{i}{\sqrt{2}}\braket{p_{c,s}}$ for the center-of-mass (c) and stretch (s) mode.}
\label{fig:trajectories}
\end{center}
\end{figure}
%

A kicking sequence with $N$ pulses displaces the Fock operators $a_{c,s}$ by a complex number $A_{c,s}$ that depends on the collective state of the ions
\beqal
a_c&\rightarrow&a_c+A_c=a_c+i(\sigma_1^z+\sigma_2^z)\alpha_c\sum_{n=1}^N e^{-i\omega_ct_n} \\
a_s&\rightarrow&a_s+A_s=a_s+i(\sigma_1^z-\sigma_2^z)\alpha_s\sum_{n=1}^N e^{-i\omega_st_n}.
\eeqal
In phase space $(\langle x_{c,s}\rangle, \langle p_{c,s}\rangle),$ the normal modes follow polygonal orbits [cf. Fig.\ \ref{fig:trajectories}]. The edges of the polygon all have uniform length $\sim\alpha_{c,s}$ and the angles between edges are determined by the arrival times of the kicks $\omega_c t_n.$ A perfect gate restores the motional state of the ion $A_c=A_s=0,$ bringing them back to their original oscillator trajectories
\beq
\label{cond}
\sum_{n=1}^N e^{i\omega t_n}=\sum_{n=1}^N e^{i\sqrt{3}\omega t_n}=0,
\eeq
and closing the orbits. Under these conditions, after a time $T$ the evolution operator becomes \cite{Garcia-Ripoll2003}
\beq
\mathcal{U}(\phi,T)=e^{-i\phi\sigma_1^z\sigma_2^z}e^{i\omega_cTa_c^{\dag}a_c}e^{i\omega_sTa_s^{\dag}a_s}.
\eeq
This is equivalent to free evolution in the trap, combined with a global phase $\phi$ that does not depend on the motional state, 
\begin{align}
\label{phi}
\phi&=\alpha_c^2\sum_{j=2}^N\sum_{k=1}^{j-1}\left[\frac{\sin(\sqrt{3}\omega (t_{j}-t_{k}))}{\sqrt{3}}-\sin(\omega(t_j-t_k))\right]\notag\\
&=:\alpha_c^2 \varphi.
\end{align}
When Eq.\ \eqref{cond} holds and the total phase satisfies
\beq
\label{phase}
\phi=\pi/4+ 2n\pi \quad n\in\mathbb{Z},
\eeq 
the combined evolution implements a controlled-phase gate on the internal state of the ions. 
The set of equations that determines the operation of the gate are solved in two steps. First, calculating the allocation positions 
$x_n=\omega t_n$, note that this allows one to re-scale the pulse arrival times $t_n$ and determines the value $\varphi$. Second, we adjust the trapping frequency to make it compatible with (\ref{phi}),
it fulfills 
\beq
\label{ws}
\omega=\frac{\hbar k^2\varphi}{16m(\pi/4+2n\pi)}.
\eeq
Note that we are allowed to overshoot the accumulated phase, exceeding the minimum value $\pi/4$ by an integer multiple $n$ of $2\pi.$ As we will see later, this allows us to fine tune the frequency, increasing $\varphi$ (i.e. more pulses) while searching for a larger overshooting factor $n.$

\subsection{Experimental setup and parameters}
\label{setup}
We propose to implement the ultra-fast two-qubit gate using $^{40}\mbox{Ca}^+$ ions confined in a Paul trap with center-of-mass frequency $\omega\in [\omega_{min},\omega_{max}]$.
The relevant internal levels of the ion are depicted in Fig.~\ref{fig:levels}a. The qubit is stored in the $4\mbox{S}_{1/2}$ and $3\mbox{D}_{5/2}$ states and we use the $4\mbox{S}_{1/2}\leftrightarrow 4\mbox{P}_{3/2}$ transition to kick the ion.

As shown in Fig.\ \ref{fig:levels}b, a single source generator produces a continuous train of pulses. A pulse picker selects pulses with discrete arrival times $t_n$ compatible with a gate protocol. The discreteness of the arrival times transforms our gate design into a combinatorial optimization problem, described in Sect.\ \ref{optimization}. Each pulse is split into two identical components by a $50/50$ beam splitter. The two pulses arrive at the ion with a relative delay $\tau,$ controlled by the relative length of the two optical paths. The ion is excited by the first pulse, which in Fig.~\ref{fig:levels}b comes from the left. By absorbing a photon, the ion acquires a momentum $+\hbar k.$ Shortly after this, a second pulse coming from the opposite direction (right in Fig.~\ref{fig:levels}b) deexcites the atom. The act of emitting a photon in the opposite direction, with momentum $-\hbar k,$ increases the momentum of the ion by $+\hbar k.$ The combined action of both pulses amounts to a very fast kick with momentum $+2\hbar k.$

To implement our phase gate, we assume a pulsed laser with these characteristics: {\itshape (i)} The laser is resonant with the ion transition, operating at a central frequency of $393.4$ nm. {\itshape (ii)} The repetition rate of the laser $R\sim 5$ GHz is much faster than the allowed trap frequencies $\omega\in 2\pi\times [78 \mbox{ kHz},2\mbox{ MHz} ]$, allowing a fine-grained control of the pulse sequences. {\itshape (iii)} The length of the pulses $\delta t$ and the delay between kicks $\tau$ are both shorter than the lifetime of the $4\mbox{P}_{3/2}$ state, $\delta{t},\tau \ll t_\gamma=6.9$ ns. This allows us to neglect spontaneous emission during the pulsed excitation and during the dark times. {\itshape (iv)} The area of the pulses is calibrated to fully transfer all probability between the $4\mbox{S}_{1/2}$ and $4\mbox{P}_{3/2}$ states, i.e. $\int_0^{\delta t}\Omega(\tau)d\tau=\pi$. Almost all requirements, except for the splitting and delay of pulses, have been demonstrated by frequency-quadrupling the light generated by a commercial laser\ \cite{Heinrich2019}.
%
%
%
\begin{figure}[t]
\begin{center}
\includegraphics[width=1.0\linewidth]{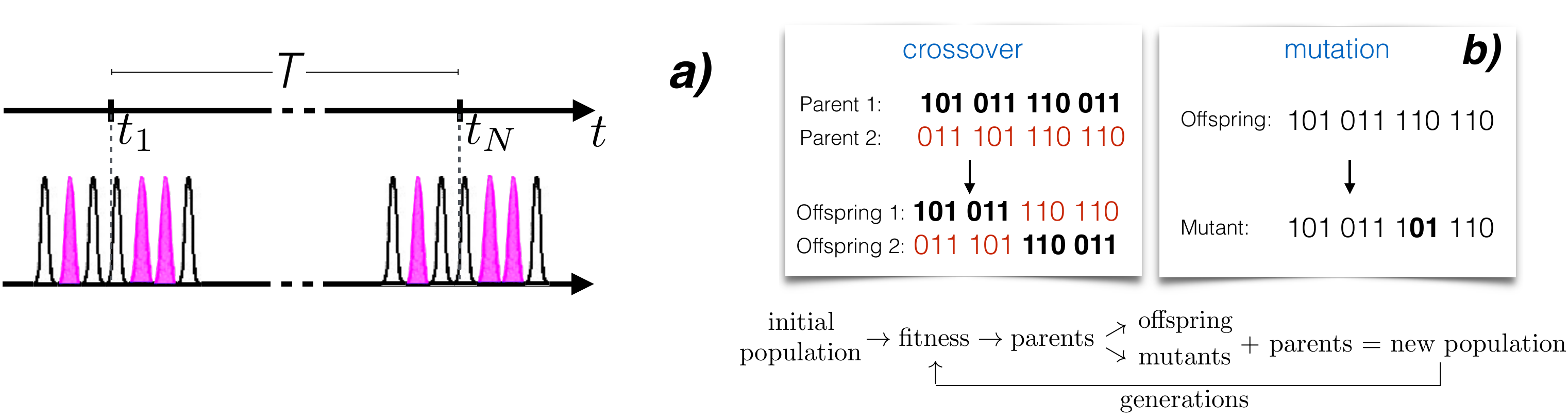}
\caption{Genetic algorithm workflow. {\itshape a)} Pulses from the equispaced sequence are allocated in $N$ discrete positions to approximate the optimal continuous delay times $t_n$. Concatenating
$M$ pulses around each optimal $t_n$ position increases the gate's phase $\phi$, keeping the same implementation time $T$. The genetic algorithm performs the optimization over $M_{\max}$ pulses around $t_n$ and selects the $M$ pulses that minimize the gate error $\epsilon$, see Eq. (\ref{error}). In the figure $M=3$ (light shadows), and $M_{max}=7$. {\itshape b)} Crossover and mutation operations performed by the genetic algorithm. In this example $N=4$, $M=2$, and $M_{\max}=3$.
}
\label{fig:optimization}
\end{center}
\end{figure}
%
%
%
\subsection{Design and optimization of a discrete control}
\label{optimization}
Section\ \ref{gate-theory} established that a control-Z gate can be implemented by a sequence of pulse pairs that satisfies Eqs.\ \eqref{cond} and\ \eqref{phase}. In this work we address the design of the pulse sequence as two consecutive tasks: (i) find a set of pulse arrival times $\{t_n\}_{n=1}^N$ that meet conditions (\ref{cond}), (ii) fine tune the trapping frequency $\omega$ so that the total acquired phase is compatible with the implementation of a CZ gate\ \eqref{phase}.

The first task decides our pulse picking strategy. This implies solving a combinatorial optimization problem, where the times $t_n = k_n \times t_R + t_1$ are spaced by integer multiples of the laser pulse period $t_R.$ In phase space, Eq.\ \eqref{cond} ensures closed polygonal trajectories [cf. Fig.\ \ref{fig:trajectories}], with angles between edges proportional to $\omega_{c,s}(t_{n+1}-t_n)$ and edge lengths proportional to $\alpha_{c,s}.$ The area enclosed by the polygons determines the geometric phase $\phi$. By adjusting the trap frequency $\omega_c,$ we tune the kick strengths $\alpha_c=\alpha_c(\omega),$ scaling the whole trajectory in phase space. This allows us to fine tune the accumulated phase\ \eqref{phase} to the desired value, modulo an irrelevant integer $n.$

The design of the pulse sequence is a hard combinatorial optimization problem, where we pick $N$ pulses out of a much longer train. To avoid the exponential complexity in this search, we find good approximate solutions using a two-stage method. The first stage is a regular minimization of the gate error\ \eqref{error} over a set of $N$ continuous arrival times $t_n\in\mathbb{R}.$ We apply a standard algorithm to minimize the gate error\ \eqref{error} over a set of $N$ variables, using $K_\text{seed}$ random initial seeds $\vec t\equiv \{t_1,t_2,\cdots,t_N\}$ of ordered times $t_{n+1}>t_n$ and $t_N\leq 2\pi/\omega$. We select a subset of $K_{\text{opt}}$ controls maximizing the phase $\phi,$ rejecting slow solutions $T>2\times 2\pi/\omega.$ In the second stage of this process, we introduce the finite repetition of the laser. We round the $K_{\text{opt}}$ continuous solutions to the nearest laser pulses, which are spaced by a multiple of $t_R=1/R.$ These discrete protocols introduce a possible timing error 
$\xi=|t_n-nt_R|.$ The gate fidelity depends on the error 
\beq
\label{error}
\epsilon=|A_c|^2+|A_s|^2,
\eeq
that we make in restoring the motional state of the ions. Instead of just minimizing each $\xi$, we minimize this global error $\epsilon$ with a genetic algorithm that fine tunes the pulse allocation. 

A genetic algorithm\ \cite{Holland1973, Holland1975} is a discrete optimizer that builds on the concept of natural selection, where solutions are iteratively improved using biologically inspired operations such as selection, crossover and mutation. In each iteration, a {\itshape population} of candidate solutions (called {\itshape individuals}) is evolved towards better solutions or {\itshape generation} based on a {\itshape fitness} function---the cost function to be optimized. On each generation, the algorithm selects a subset of individuals that maximize the fitness. These so called {\itshape parents} merge and mutate, giving rise to new solutions, the {\itshape offspring} that form the next generation. This process of selection and reproduction is repeated until the fitness reaches the desired optimal value, selected by a user-defined tolerance, or until the maximum number of generations is reached.

To bring our problem into this form, we take the $N$ continous times $t_n$ and find out the $M_{\max}$ closest pulses within the sequence created by the laser [cf. Fig.\ \ref{fig:optimization}a]. We then encode a solution as a \textit{chromosome} with $N\times M_{\max}$ \textit{genes}. Each gene is a bit that becomes $1$ when the corresponding pulse is selected [cf. Fig.\ \ref{fig:optimization}b]. Our initial population is formed by $K_\text{ind}$ individuals, each with $N\times M$ active genes, indicating that we have $N$ groups of $M$ pulses around the times $t_n.$ From this pool, we select the $K_p$ individuals exhibiting the best value of the fitness function (\ref{error}). Parents mate in pairs and each child receives part of its chromosome from the first parent and the rest from the second. In our algorithm this proportion is $50/50$ made at the middle of each parent chromosome, see Fig.\ \ref{fig:optimization}b. If a child improves the fitness function it joins the parents to constitute the new population for the next generation. If not, a mutation is produced creating random variations in the chromosome. To preserve the total number of $N\times M$ pulses, we randomly swap the values of two genes from a $M_{max}$ sequence placed around one of the times $t_i,$ see Fig. \ref{fig:optimization}b. These mutants join the new population, irrespective of their value of the fitness function, and the whole process is repeated. This workflow, sketched in Fig.\ \ref{fig:optimization}, is repeated over $K_{\text{ite}}$ generations. At the end, we select the state that produces the best value of the fitness function, thereby minimizing the error Eq.\ \eqref{error}.

\section{Results}
\label{results}

\begin{figure}[t]
\begin{center}
\includegraphics[width=1.0\linewidth]{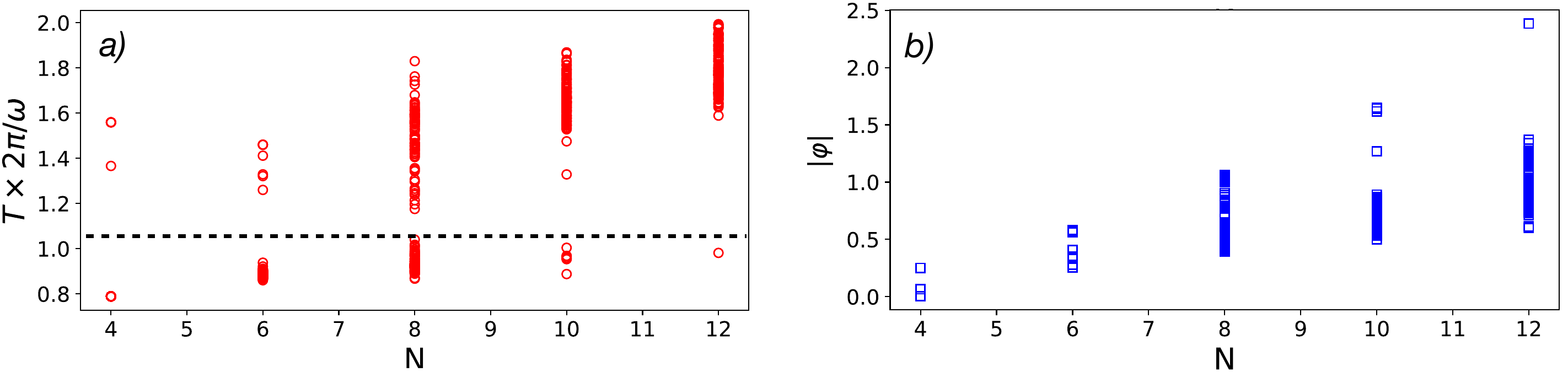}
\caption{Partial gate optimization assuming full control over pulse arrival times. {\itshape a)} Duration of pulse sequences satisfying Eq.\ \eqref{cond} as a function of the number of applied pulses $N$. The figure only shows solutions with a duration $T<2\times 2\pi/\omega$. The fastest combination within each $N$-sequence fulfills $T\lesssim 1.055\times 2\pi/\omega$ independently of $N$ (black dashed line). {\itshape b)} Corresponding phase as a function of $N$.}
\label{assym}
\end{center}
\end{figure}

As mentioned above, our simulations consider a scenario where the direction of the kicks is fixed. This happens when a single pulse picker is connected to an interferometric setup, creating pairs of pulses all arriving with the same relative delay [cf. Fig.\ \ref{fig:levels}]---e.g. the left pulse always excites the ion and the right pulse immediately de-excites it, setting $z=+1.$ Scenarios where both the relative direction and the Lamb-Dicke parameter are tuned have been considered before\ \cite{Garcia-Ripoll2003, Duan2004, Bentley2013, Gale2020} leading to different degrees of controllability and thus to different gate times. Here we will show that, despite our experimentally-motivated constraints\ \cite{Heinrich2019}, it is possible to implement CZ gates in a time shorter than the trap period $T<2\pi/\omega.$

\begin{figure}[t]
\begin{center}
\includegraphics[width=1.0\linewidth]{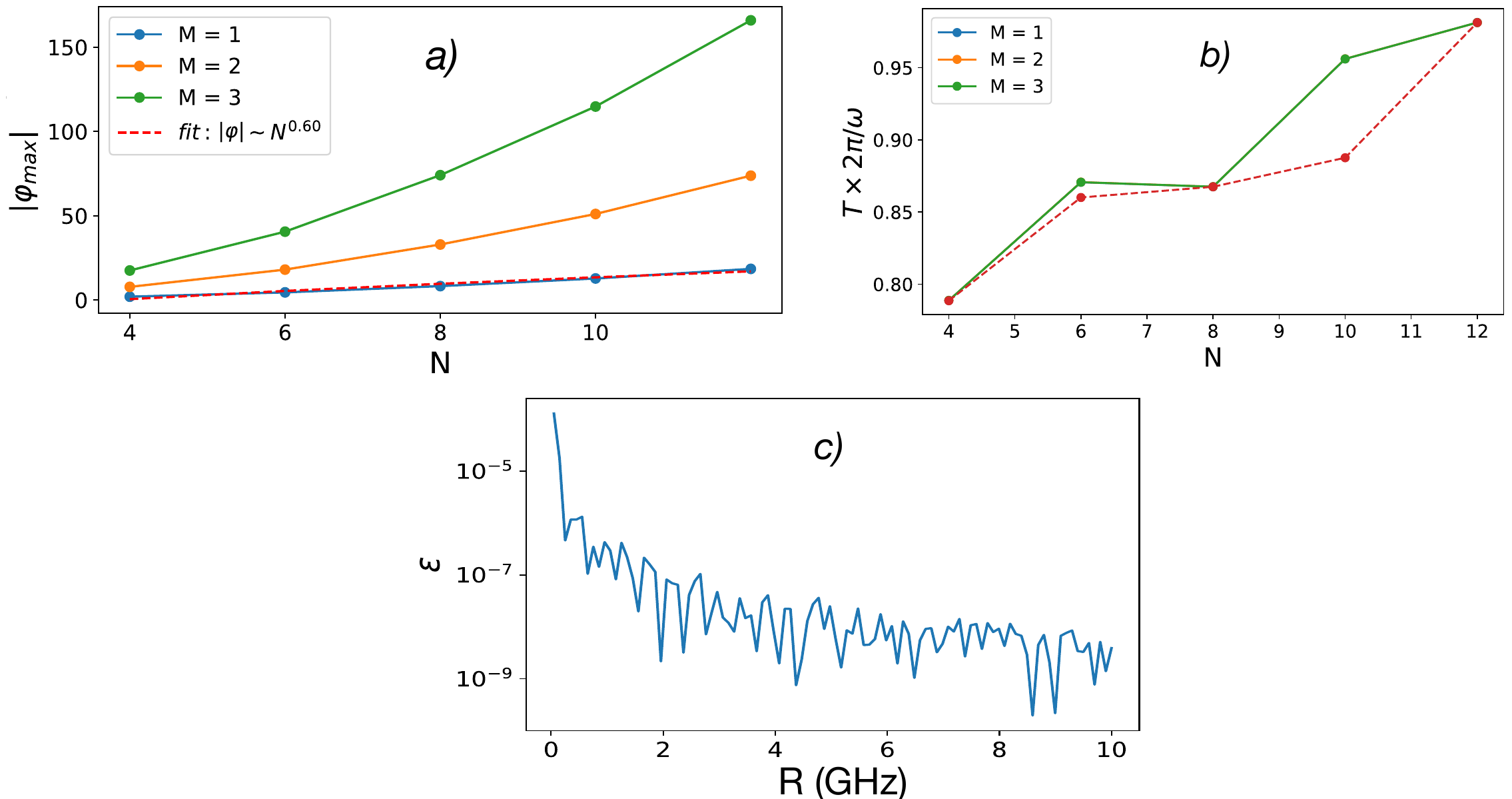}
\caption{Optimal gates for discrete pulse arrival times. {\itshape a)} Phase vs. number of batches $N$ and number of pulses per batch $M,$ for $R=5000\omega.$ When $M=1,$ maximum phase (blue solid line) grows as $|\varphi| \sim N^{0.6}$ (red-dashed line). Clustering $M=1,2,3$ (blue, orange, green) of pulses around the optimal $t_n$ times allow us to increase the phase.
{\itshape b)} The total gate duration is insensitive to the number of pulses per batch, and approaches the shortest gate for continuously varying pulse arrival times (dashed red).
{\itshape c)} Error (\ref{error}) of the phase gate ($N=6, M=1$) as a function of the repetition rate $R$.
}
\label{discrete}
\end{center}
\end{figure}
%
%
%
%
%
Before illustrating the final protocols, Fig.\ \ref{assym} shows the intermediate results obtained when solving the commensurability equations\ \eqref{cond} with continuous variables $\{t_n\}_{n=1}^N.$ Note how for a fixed number of pulses $N$ there exist multiple schemes that restore the motional state of the ions and implement a control phase gate. Out of those combinations we select those that maximize the ratio $\varphi=|\phi/\alpha_c^2|,$ and feed them to the genetic algorithm to create discrete pulse sequences. Note that the two-qubit phase depends on $\alpha_c$ and therefore on the trap frequency $\omega_c.$ The preselection of continuous protocols with large $\varphi$ provides a broader choice of pulse sequences and frequencies (\ref{ws}) that satisfy both the experimental restriction $\omega\in[\omega_{min},\omega_{max}]$ and the phase relation Eq.\ \eqref{phase}, with either $n=0$ or $n\neq 0$ (\textit{overshooting}).

The accumulated phase grows with the number of pulses in the discrete protocol as $|\phi|\propto N^{0.6},$ [cf. Fig.\ \ref{discrete}a], while the duration of the gate remains below $T\lesssim 1.055\times 2\pi/\omega$ and is close to the sequences minimizing the gate time $T$ [cf. Figs.\ \ref{assym}a and \ref{discrete}b]. The error introduced by the finite repetition rate is also negligible, Fig.\ \ref{discrete}c shows the theoretical error for one protocol consisting of $N=6$ pulses. A laser with a repetition rate $R\gtrsim 1$ GHz already produces an ultra-fast two-qubit gate with fidelity above $99.999\%.$

\begin{figure}[t]
\begin{center}
\includegraphics[width=1.0\linewidth]{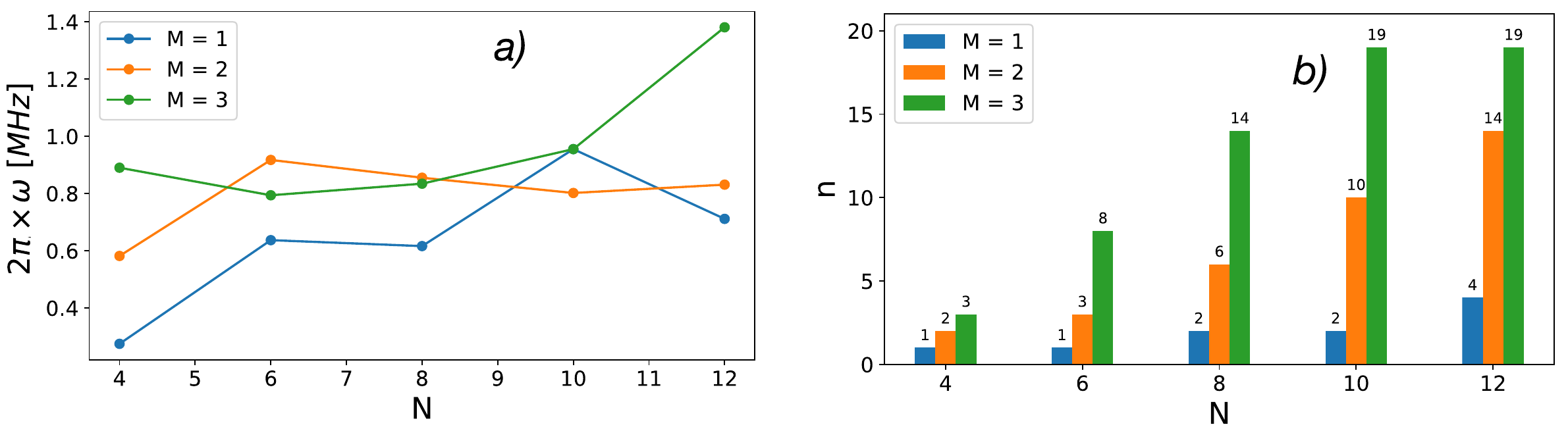}
\caption{{\itshape a)} Trapping frequencies associated with each phase gate depicted in Fig.\ \ref{discrete}. The frequencies are tuned such the frequency fulfilling (\ref{phase}) is closest to the working trapping frequency $\omega=2\pi\times 0.82\ \mbox{MHz}$ \cite{Heinrich2019}.
{\itshape b)} Overshoot value of the phase $\phi=\pi/4$ associated with the trapping frequencies showed in {\itshape b)} when implementing the gate.
}
\label{fig:phase}
\end{center}
\end{figure}

As shown in Fig.\ \ref{discrete}b, a short sequence with $N=4$ pulses produces very fast gates $T<2\pi/\omega,$ but with a small acquired phase. We may increase the accumulated $\varphi$, concentrating $M$ pulses around each of the $N$ kicking times [cf. Fig.\ \ref{fig:optimization}a]. This maintains the shape of the orbits, scaling the edges by a factor of $M$ [cf. Fig.\ \ref{discrete}a]. As shown in Figs.\ \ref{discrete}a-b, the duration of the gate is preserved and the accumulated phase grows with the area as $\varphi\propto M^2.$ Note that, since the phase increases in discrete steps, we still need to fine tune the trap frequency to match the desired CZ. Figures\ \ref{fig:phase}a and \ref{fig:phase}b show that this is possible for realistic trapping frequencies\ \cite{Heinrich2019}, using different multiplication factors $M.$ Figure\ \ref{fig:phase}a shows the frequencies (\ref{ws}) that implement a CZ gate and which are closest to the desired value $\omega\sim 2\pi\times 0.82$ MHz. As $\varphi$ grows with both $N$ and $M$, Fig.\ \ref{fig:phase}b shows that the specific frequency is achievable compensating the phase with a large overshooting factor $n$.
%
%
%
%
%
%
%
%
%
%
%
%
\section{Estimation of errors}
\label{errors}
We have presented a route for the implementation of ultra-fast $T<2\pi/\omega$ quantum gates using a train of laser pulses that are resonant with the transition frequency of a trapped ion. In these protocols, the motional state of the ion is almost perfectly restored with a high-fidelity $\epsilon \sim 10^{-9}-10^{-7}$ using source generators with a constant repetition rate $R\sim 5$ GHz.

When implementing these protocols, actual experiments will suffer from imperfections in the control of the ion, due to spontaneous emission during the time that the ion remains in the excited state  $4P_{3/2}$ (i.e., during pulses and waiting time), and due to intensity fluctuations in the pulses.

A trivial model to quantify the spontaneous emission errors, giving an upper bound on them, is to write a density matrix
\beq
\rho = (1-P_{err}) |\psi\rangle\langle\psi| + P_{err} |g\rangle\langle g|
\eeq
where $|g\rangle$ is a fictitious state accumulating the probability that an error took place. The fidelity is given by $P_{err}F_0,$ where $F_0$ is the fidelity of the gate implemented by ideal kicks. In this model, $P_{err}$ feeds from spontaneous emission effects: we assume that whenever the emission takes place, the experiment must be repeated. The probability that the ion is in an excited state $|e\rangle$ is
\beq
\frac{dP_{ok}}{dt} =- \gamma |\langle e|\psi(t)\rangle|^2 (1-P_{err}),
\eeq
with $P_{ok}+P_{err}=1.$ The decay rate $\gamma=1/t_\gamma$ is inversely proportional to the lifetime $t_\gamma$. The solution to this problem is
\beqa
\epsilon_{\gamma}& =& 1 - P_{ok}(T), \\
P_{ok}(T) &=& \exp\bigg(-\gamma \int_0^T  |\langle e|\psi(t)\rangle|^2 dt\bigg) P_{ok}(0). \nonumber
\eeqa
In a very crude scenario, we upper bound the error probability, assuming that the ion is in the excited state from the beginning of the exciting pulse, to the end of the following, that is $T_e\simeq \delta t + \tau,$
\beq
\epsilon_{\gamma} = 1 - \exp(-\gamma T_e) \simeq \gamma T_e.
\eeq
In the experimental setup from Fig. \ref{fig:levels}b, the waiting time $\tau$ between counter-propagating pulses is controlled by the relative length
of the optical paths. The minimum separation is given by the pulse duration, $\tau\gtrsim\delta t$ to avoid interference.
In our system, the excited state $4P_{3/2}$ has a lifetime $t_\gamma=6.9$ ns and $T_e\simeq 1$ ps\ \cite{Heinrich2019, Hussain2020} leading to errors $\epsilon_{\gamma}\sim\mathcal{O}(\delta t/t_\gamma)\simeq 1.4\cdot 10^{-4}$.
For a sequence containing $N$ kicks, the infidelity of the gate is approximately $\epsilon_{\gamma}^{gate} = 1-(1-\epsilon_{\gamma})^N\sim\mathcal{O}(N\delta t/t_\gamma)$.

We can also quantify the errors $\epsilon_A$ due to fluctuations in the $\pi-$pulses. For a general pulse shape $\theta=\int_0^{\delta t}\Omega(\tau)d\tau$ the unitary generated
by the interaction Hamiltonian (\ref{Hi}) is
\beq
\hat U_k=\bigg(c-is\hat\sigma_1^xe^{ikx_1\hat\sigma_1^z}\bigg) \bigg(c-is\hat\sigma_2^xe^{ikx_2\hat\sigma_2^z}\bigg)
\eeq
with $c=\cos(\theta/2)$ and $s=\sin(\theta/2)$. A perfect $\pi-$pulse, i.e. $\theta=\pi$, generates the unitary
$\hat U_{kick}=-\hat\sigma_1^x\hat\sigma_2^xe^{ik(x_1\hat\sigma_1^z+x_2\hat\sigma_2^2)}$. In order to quantify the errors due to area fluctuations when combining two
counter-propagating pulses $\hat U_{pair}=\hat U_k\hat U_{-k}$ we consider small fluctuations $\pi+\Delta\theta=\int_0^{\delta t}\Omega(\tau)d\tau$ (with $\Delta\theta\rightarrow 0$) in the pulse area.
Retaining the first order terms in $\Delta \theta$ an imperfect pair of counter-propagating pulses generates the transformation
\beq
\hat U_{pair}=(1-\Delta\theta^2/2)\hat U_0-\Delta\theta\hat U_e^1-\Delta\theta^2\hat U_e^2+\mathcal{O}(\Delta\theta^3)
\eeq
with $\hat U_0=e^{-2ik(x_1\hat\sigma_1^z+x_2\hat\sigma_2^z)}$ the optimal unitary generated by two perfect counter-propagating $\hat U_{kick}$ pulses, and
$\hat U_e^1=i(\sigma_1^x\cos(kx_1)e^{-2ik\hat\sigma_2^zx_2}+\sigma_2^x\cos(kx_2)e^{-2ik\hat\sigma_2^zx_1})$ 
and $\hat U_e^2=\cos(kx_1)\cos(kx_2)\hat\sigma_1^x\hat\sigma_2^x+(e^{ikx_1\hat\sigma_1^z}+e^{ikx_z\hat\sigma_2^z})/4$
accounting for unrestored and incorrect motional dynamics. The total unitary of a gate 
can be approximated by the product of $N$ pairs 
\beq
\label{area}
\hat U_{gate}\approx (1-\Delta\theta^2N/2)\hat U_{N}-N\Delta\theta\hat U_{err}
\eeq
with $\hat U_N=\hat U_0^N$ 
and collecting all the errant dynamics in $\hat U_{err}$ that it is assumed orthogonal to the ideal unitary $\hat U_0$. This is a conservative approximation that neglects terms that result in an incorrect motional state, but includes those that correctly restore the internal state \cite{Gale2020}. For any initial state $|\psi\rangle$ of the computational basis we can compare the dynamics of the optimal gate $\hat U_{opt}$ with the
one generated by $\hat U_{gate}$. To this end we estimate the fidelity   
\beq
F=|\langle\psi|\hat U_{opt}^{\dag}\hat U_{gate}|\psi\rangle|^2 = (1-N\epsilon_A+N^2\epsilon_A^2/4)F_0,
\eeq
with $\epsilon_A=\Delta\theta^2.$ The magnitude of the fluctuations $\epsilon_A$ depends on the specific characteristics of the laser pulses. In real setups with picosecond pulses\ \cite{Heinrich2019, Campbell2010} these fluctuations are found to induce errors of around $\epsilon_A\propto \Delta I/I\sim 10^{-3}$. However, these intensity fluctuations can be reduced experimentally, using methods such as adiabatic rapid passages with chirped laser pulses\ \cite{Malinovsky2001, Wunderlich2005, Heinrich2019b}.

%
%
%
%
%
%
%
%
%
\section{Outlook}
\label{outlook}

Our analysis shows that it is possible to engineer ultra-fast gates $T<2\pi/\omega$, using pulse picking strategies for an experimentally relevant setup\ \cite{Heinrich2019, Hussain2020}.
Current two-qubit M{\o}lmer-S{\o}rensen gate operations require a duration of $\bar T\sim 40$ $\mu$s for entangling two qubits at a trapping frequency $\omega\simeq 2\pi\times 1.4$ MHz\ \cite{Bermudez2017}. Compared to these numbers, our scheme can provide a speedup factor $\bar T/T> 50$, for a conservative gate duration $T\sim 2\pi/\omega.$

Our investigation leaves some open questions, to be addressed in later works. The first one concerns the robustness of the protocol with respect to intensity fluctuations and spontaneous emission. Both problems may be overcome if we use STIRAP techniques \cite{Bergmann1998, Vitanov2017, Shapiro2007}, to induce excitation between the $4\text{S}_{1/2}$ and a metastable state, such as $3\text{D}_{5/2}$ or $3\text{D}_{3/2}$. Experimentally, ${}^{40}$Ca$^+$ ions have been robustly manipulated using such techniques \cite{Sorensen2006, Moller2007, Timoney2011}. For our proposal, we could detune the pulsed laser exciting the $4\text{S}_{1/2}\to 4\text{P}_{3/2}$ transition and combine it with another pulse connecting the $4\text{P}_{3/2}\leftrightarrow 3\text{D}_{5/2}$ states. These improvements can be supplemented with pulse shaping techniques\ \cite{Palao2002, Romero-Isart2007, Doria2011}, to minimize the AC Stark-shifts and dephasing associated with high-intensity pulses.

A second, more pressing question, concerns the parallelizability and scalability of our pulsed schemes. Recent works have addressed theoretically\ \cite{Garcia-Ripoll2005, mehdi2020a, mehdi2020b} and demonstrated experimentally\ \cite{figgatt2019,lu2019} the simultaneous implementation of arbitrary two-qubit gates among a subset or all pairs of $K$ ions in a trap. We can use our two-step protocol to perform this task with significant speed ups. As in this work, the first step is a continuous optimization of the desired gate operation, subject to the now $2K$ dynamical constraints\ \cite{Garcia-Ripoll2005}. The resulting pulsed protocol is fine tuned with our genetic algorithm, to match the repetition rate of the laser. The process has an increased optimization cost, but the multi-qubit gates do not seem to take longer than the two-qubit ones\ \cite{Garcia-Ripoll2005}.

Current ion trap quantum computers are able to run programs with up to several hundred one and two-qubit operations\ \cite{Martinez2016}. We expect that these methods and subsequent improvements ion trap quantum computers will be able to improve at least one, if not two orders of magnitude, leading to an increased quantum volume in NISQ devices. Moreover, the estimated ideal gate fidelities are compatible with existing error thresholds\  \cite{Knill2005}, which makes these methods a promising alternative for implementing fault-tolerant computation schemes\ \cite{Bermudez2017}.

\ack
We acknowledge support from Project PGC2018-094792-B-I00 (MCIU/AEI/FEDER,UE), CSIC Research Platform PTI-001, and CAM/FEDER Project No. S2018/TCS-4342 (QUITEMAD-CM). Authors also acknowledge support by the Institut f\"ur Quanteninformation GmbH. \\

%
%
\bibliographystyle{iopart-num}
\bibliography{fast_ion.bbl}

\end{document}